# Polar charge induced self-assembly: An electric effect that causes non-isotropic nanorod growth in wurtzite semiconductors

Yury Turkulets, and Ilan Shalish*

*Ben Gurion University of the Negev, Beer Sheva 8410501, Israel*

Crystals grow by gathering and bonding of atoms to form an ordered structure. Typically, the growth is equally probable in all crystalline directions, but sometimes, it is not, as is the case of nanowire growth. Nanowire growth is explained, in most cases, by the presence of liquid metal droplets that mediate between an incoming flux of atoms and a substrate or an existing crystal nucleus, while defining the lateral dimension. Here, we report and explain a previously unknown mode of non-isotropic crystal growth observed in two wurtzite semiconductors, InN and ZnO. Being of polar structure, wurtzite crystals possess a built-in internal electric field. Thermally-excited charges screen the built-in electric field during growth in a non-uniform, yet symmetric, manner, causing the formation of symmetric domains of inverted polarity. These domains limit the lateral expansion of the crystal, inducing a fiber growth mode. The mechanism described here elucidates previously unexplained phenomena in the growth of group III-nitrides on sapphire, emphasizing the need to consider the effects of built-in electric fields in the growth of polar semiconductors.

While the interest in polar semiconductors is constantly rising, understanding of their physics has been lagging behind. As we show here, the inherent internal electric fields may have unique anisotropic effect on the growth of polar semiconductors crystals. Anisotropic crystal growth has been a subject of major interest by its own virtue. One useful result of it is nanowires – nanometer scale crystalline fibers that have been firing the imagination of the worldwide scientific community as a possible building block for nanotechnology.[1] Although nanowire growth has been intensively studied,[2] not all is known yet about the reasons, for which crystals grow as fibers. The first explanation, suggested by Frank, described them as evolving around a screw dislocation.[3] Infrequently encountered in fibers, screw dislocations invoked much controversy and debate, until the model was joined by the now widely accepted vapor-liquid-solid model, proposed by Wagner and Ellis of Bell Labs in the 1960s.[4] Their model suggested that a droplet of liquid metal mediates the growth, restricting the lateral size of the fiber to the droplet contact area. Successfully tested, this model has become the basis of the main technique for nanowire growth today.[5] Yet, nanowires often grow without an intentionally added catalyst.[6] Most of these cases are classified as "self-catalysis", i.e., catalysis by the metal ingredient in the nanowire compound.[7,8] Such non-catalyzed growth mode is often considered cleaner, but at the same time, it is also considered more challenging in terms of the control of wire diameter.[9] Interestingly, the resulting nanowires are sometimes uniform and reproducible in size and shape.[10] Could there be a mechanism that, in the true and complete absence of a catalyst, drives crystals to grow in fiber form?

Along with the extensive research of nanowires, various modifications of the vapor-liquid-solid model have been proposed, but they all required the involvement of a catalyst material (see Kolasinski *et al*. and references therein).[11] During our initial attempts to grow InN nanowires, our findings suggested that the fibrous crystal growth we obtained followed a mechanism that differed from anything known before. Later, we observed the same in ZnO as well. Here, we describe our observations and propose a mechanistic model that provides a step-by-step physical explanation for a nanowire growth mode that does not require a catalyst.

InN marks the lower bandgap limit of the nitride semiconductor family.[12] The members of this family do not occur in nature and are typically grown epitaxially on sapphire (an insulator) or silicon carbide (6H-SiC, a semiconductor). Solid solutions of InN and GaN and/or AlN are used to engineer quantum structures with varying bandgap for photonic applications that could potentially span a wide photonic spectrum, from the infrared to the ultraviolet (1700 to 200 nm).[13]

Under the conditions described here, InN grows on c-plane sapphire as submicron fibers (nanorods), in a unique, symmetric, and uniform mode of growth that does not seem to match any of the existing models for nanowire growth. To explain the mechanism of this phenomenon, we propose a model, in which *electric-charge-driven* symmetric and reproducible polarity inversion processes predictably and accurately determine a limit to the nucleus lateral expansion to produce outstandingly narrow distribution of rod diameters (Fig. 1).

Grown on c-plane sapphire substrates, the InN rods showed epitaxial relation to the substrate (Fig. 1A and 1B). This was despite the lattice constant differences (the 'a' lattice constant is 0.4785 nm for sapphire and 0.3545 nm for InN) The experimental details of the growth as well as basic material characterizations have already been given elsewhere.[14]

The rods were randomly spaced (~0.3 rods/$\mu m^2$) and appeared to have rather narrow distribution of diameter (373 ± 22 nm) and length (1723 ± 58 nm). The correlation between diameter and length, as measured in SEM images with pixel size of ~10 nm, was 0.53. Rod lengths were roughly equal to 300 'c' lattice constants, while their diameters were roughly equal to 100 'a' lattice constants. Since the structure is 2H, the period in the c-axis direction was actually 2 times that of the 'c' lattice constant (stacking order of ababab… - one 'c' lattice parameter for 'a' and another for 'b'). Hence, the height was about 1.5 times the diameter in terms of the 2H-InN polytype unit cell, i.e., the average crystallite was about 150 unit cells long and 100 unit cells in diameter. The rods appeared to grow in perpendicular orientation to the substrate, ending with hexagonal pyramidal tips. The main peaks observed in 2θ-ω symmetric X-ray diffraction were clearly identified as $Al_2O_3$(0006), InN(0002), and InN(0004), which confirmed what is already suggested in the SEM image, i.e., the rods adopt the c-orientation of their sapphire substrate. The InN(0002) peak was combined with a minor InN(10-11) peak diffracted from the sloped sides of the pyramidal tip.

Head-on SEM images show the hexagonal cross-section of the rod tops (Fig. 1B). However, close inspection of the crystallite sides (Fig. 2A) reveals a more complicated structure, comprising what appears to be two interwoven hexagonal phases. Figure 2B is a schematic depiction of the crystal, in which the two phases are shaded differently. The apparent cross sections at various heights are shown to the left, while a drawing of the bottom phase alone appears to the right. One phase occupied most of the rod's cross-section at its base, while the other dominated at the tip. The two phases appeared to share the volume of the rod in perfect symmetry that was accurately replicated in virtually all the rods we examined (within the practical limit of the number of rods that could be feasibly examined).

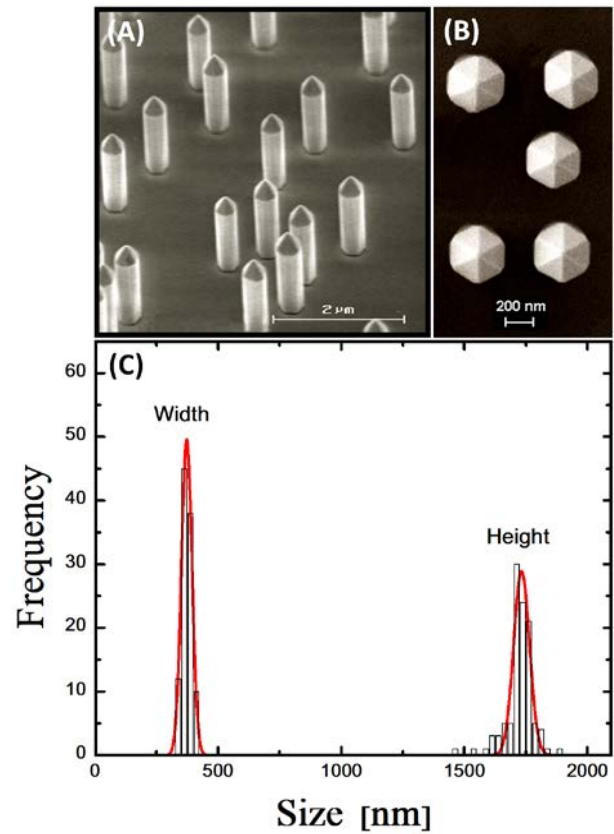

**Figure 1** - Scanning electron microscope (SEM) images of InN nano-rods on sapphire: (A) 30°, and (B) 90° (head-on view). (C) Histogram of widths and heights obtained from 100 rods in SEM images. A shows the uniform appearance of the rods, while B shows that the rods are aligned with each other, suggesting epitaxial relations with the substrate. C shows the narrow distributions of rod dimensions.

Rod growth seemed to begin with one phase, while the other phase only occupied the six corners of the hexagonal base. As the growth proceeded, the corner phases expanded inward toward the rod's center, and upward from all six corners until they occupied the entire cross section. Following this point, the rods ended their growth in a pyramidal shaped tip. Selected area diffractions did not show more than a single crystal.[14] Therefore, the only way to explain the observation of two phases is that they are polarity-inverted domains, which are indistinguishable by this diffraction. To test this hypothesis, we used converged beam electron diffraction (CBED). A single rod was first encapsulated in Pt and attached to a tungsten tip in a focused ion beam (FIB) lift-out transmission electron microscopy (TEM) grid, and then thinned down using a Ga ion beam to a thickness of about 100 nm. Figure 2C shows a TEM image of a nano-rod attached to a tungsten tip. Figure 2D is a schematic illustration of the same. CBED patterns were then acquired from the middle of the base and from one side of the pyramidal tip. A comparison of measured and simulated CBED patterns confirms that at the base, the rod grew in the (0002) direction, while at the tip, it grew in the $(000\bar{2})$ direction (Fig. 2E).

The above observations elicit two questions. First, what limits the lateral growth to a specific size that is rather accurately reproduced? And second, what mechanism can underlie such a symmetrically interwoven polarity inverted domain structure? In what follows, we attempt to answer the above questions using a single mechanism.

The In-N bond is of ionic nature, its crystal lacks inversion symmetry, and the ratio of its lattice parameters is smaller than that for perfectly hexagonally close-packed atoms (c:a=1.613 for InN and 1.6333 for hcp). Together, these features produce a net dipole moment along the c-axis of the unit cell. Although these dipole moments cancel out among adjacent dipoles in the bulk, uncompensated opposite polarization charges remain on the two polar faces. The resulting energy band diagrams are shown in Fig. 3A. Ab-initio calculations showed a rather large spontaneous polarization in InN of $\sigma_{SP} = -3.2 \cdot 10^{-6} \frac{Cb}{cm^2}$ which is equivalent to a sheet carrier concentration of $2 \cdot 10^{13}\ cm^{-2}$ of positive charge at the N-polar face and an equal concentration of negative charge on the In-polar face.[15] The resulting built-in field along the polar axis of the crystal is modified during the high temperature growth by the pyroelectric effect. The difference between the 'a' lattice constants of sapphire and InN adds a piezoelectric component to the built-in field that augments the spontaneous polarization at the early stages of the growth in the proximity of the substrate interface. Due to controversy on calculated values of the spontaneous polarization**Error! Bookmark not defined.** and lack of data on the pyroelectric effect in InN,[16] it is not possible at this time to evaluate the net polar built-in field. However, for the purpose of our model, we only need to assume that the combined effect of these three contributions is a non-zero built-in electric field. We will also assume that the net field points in the (0002) direction. If, in fact, it actually points otherwise, i.e., in the $(000\bar{2})$ direction, the sign of the charges at each polar face will change, but this would not affect the model. Due to the small bandgap (0.65 eV)[17] and the high growth temperature (~550 °C), electron-hole pairs are generated thermally during the growth. The field induced by the polar charges, situated on the polar faces, attracts electrons to the N-polar face and holes to the In-polar face. This is shown in an energy band diagram in Fig 3A (we follow a similar band diagram proposed for GaN[18]).

Growth begins from a nucleus that initially expands horizontally. Electron-hole pairs separated by the polar built-in field are swept to the polar faces. However, due to the small diameter of the crystal, carriers of the same charge experience mutual coulomb repulsion that sweeps them away from each other as far as they can go (illustrated in Fig. 3B). In a hexagonally-shaped InN mesa, they divide equally among the six corners. As the growth starts along the c-axis, the top of the crystal mesa is the In-polar face, and the holes that are swept upward crowd at the six top corners. At the same time, electrons do the same on the N-polar side (interfacing the substrate). Thus, at the center of the hexagonal mesa, where there are no free carriers, the electric field is only a superposition of the spontaneous polarization field, $F_{SP}$, the pyroelectric field, $F_{PZ}$, and the piezoelectric effect, $F_{PE}$. However at the corners, the crowding mobile charges partially compensate the polar charges, effectively inducing an opposite field, $F_{FC}$, that weakens the polar field

$$\sum F = F_{SP} + F_{PZ} + F_{PE} - F_{FC}$$

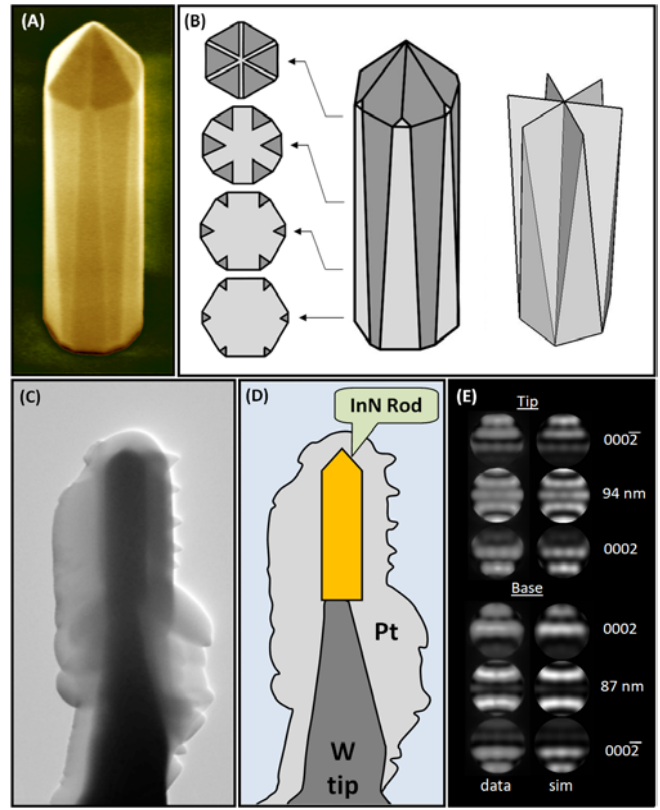

**Figure 2** – (A) SEM close up image of a single rod, false colored to enhance the details. (B) Schematic illustration of a rod emphasizing the two phases, showing cross sections at different rod heights on the left, and the non-inverted phase alone on the right. (C) TEM image of a single nano-rod attached to a tungsten tip on a FIB TEM grid. The rod is coated with Pt and thinned down to about 90 nm. (D) Schematic map of Fig. C showing its different parts. (E) Left column: Two converged beam electron diffraction (CBED) patterns obtained from the center of the rod base (marked as point #2) and from one side of the rod tip (marked as point #1). Right column: Simulated CBED patterns are shown for comparison to the right of each measured pattern.

(illustrated in Fig. 3C). As the hexagonal mesa expands, its volume increases and (per the same thermal generation rate) the total number of generated electron-hole pairs increases. At a certain width, the mobile charges at the corners reach a large enough density to totally cancel the polar charge (note that this happens only at the corners). Subsequent expansion from that point and on contributes additional charge to create a greater opposite field at the corners, until eventually, the net field at the corners is inverted.

As long as the electric field is pointing up, the added columns maintain the crystal's original polarity. At some point, however, the electric field at the corners flips, and from then on, maintaining the growth in the original polarity becomes increasingly difficult. Eventually, the electric stress becomes so high that the only way to relieve it is to flip over the polarity, placing an inverted dipole at the corner. Once this happens, the corners become effectively "locked", thereby precluding further horizontal expansion. This is because at the vicinity of the inverted corners, the polar electric fields are less defined, and new material is lacking a definite electrical guidance. Since the thermal generation rate is roughly the same in all nuclei, they all become locked for further horizontal expansion at about the same width (volume). This mechanism explains the narrow diameter distribution of the rods. More importantly, it explains why the growth forms nanorods rather than continuous layers. We note that this "locking" is not absolute as expansion and sideway growth of the facets appears to continue at a slower pace, and this seems to be the reason why the inverted phase does not appear to reduce to zero width at the substrate height level. This is more pronounced in ZnO that will be shown later.

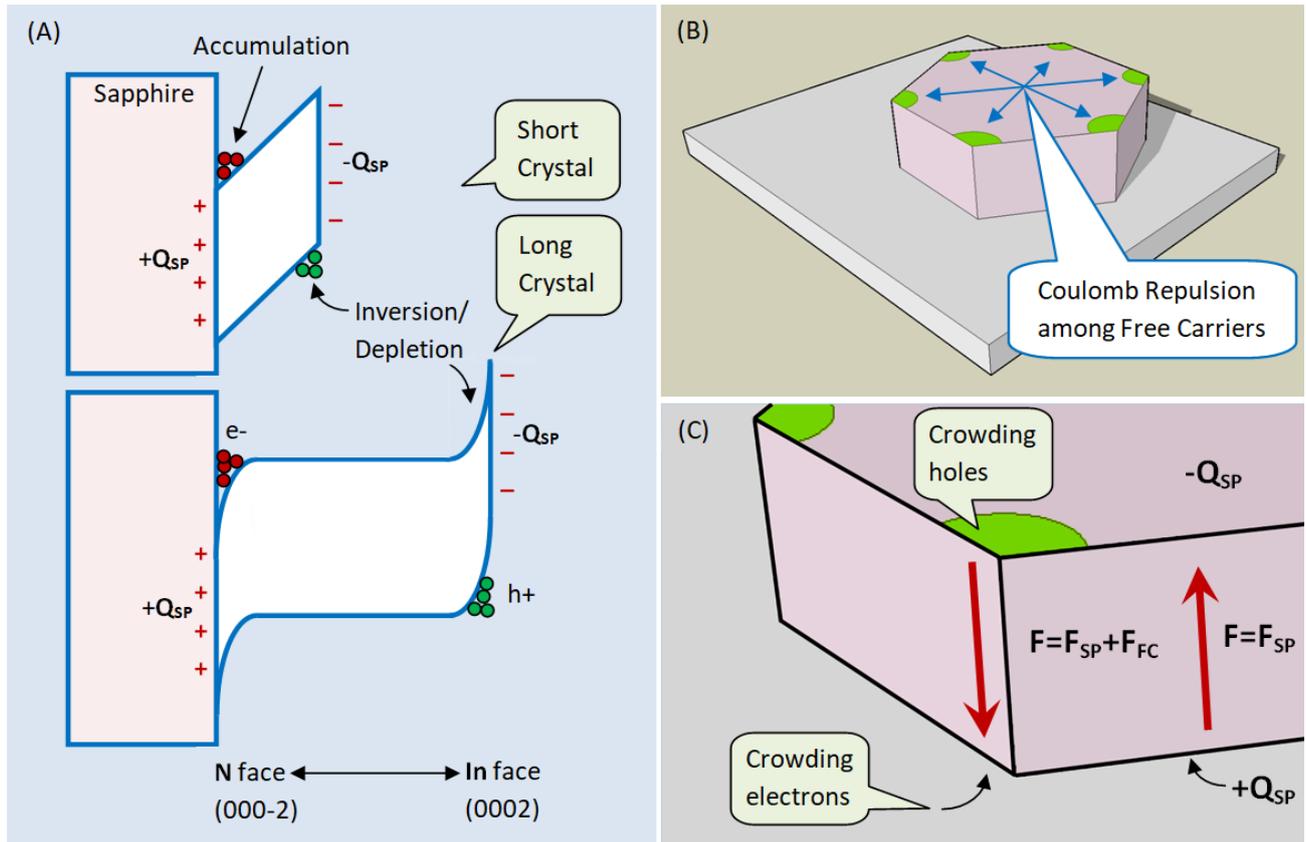

**Figure 3** – (A) Energy band diagrams showing the effect of the polar charges inducing built-in fields that result in surface accumulation on the In-polar face and surface inversion/depletion on the N-polar face, following the model of Harris et al for GaN.[18] At the beginning of the growth the rod is very short and is fully depleted as shown in the top diagram. As the rod continues to grow, it becomes long enough such that the built-in fields are eventually limited to the surface regions. (B) schematic illustration of a hexagonal, mesa-shaped, nucleus at the beginning of the growth process showing that mobile charges on the top surface crowd into to the six hexagon corners. Similar crowding takes place at the bottom face of the mesa with opposite sign charges. (C) Zoomed in view of B showing the electric field at a corner, where the direction of its electric field flipped relative to the nearby electric field situated some distance from the corner.

Once the mesa becomes "locked" for horizontal expansion, it proceeds to grow upward. As the growth proceeds upward, each additional layer increases the volume and, consequently, increases the number of generated electron-hole pairs. This increases the number of free carriers crowding at each corner, causing them to gradually spread toward the center of the mesa. As a result, the inverted polarity phase at the corner gradually expands toward the center as the height of the rod increases. This explains why the inverted phase at the corners is observed to expand inward until, at a certain height, all six inverted phases from the six corners meet each other at the center, thereby eliminating further growth of the original, non-inverted, phase. At this point of the crystal growth, one could expect the same process of corner inversion to start over. However at this point, the volumes of the two phases become identical. One phase separates electron-hole pairs sending the *holes* upward, while the other phase sends the *electrons* upward. As a result, the net mobile charge at the top is small and there is not enough of it to cause another corner-inversion.

The pyramidal tip formation is not necessarily a part of the present model. Having been observed in both polar and non-polar materials, pyramidal growth does not seem to always require charges and electric fields.[19] It has been studied and explained in several models.[20,21] In our case, the pyramidal tip is reminiscent of the pyramidal V-pits commonly observed in InGaN, the sloped sides of which are also $(10\bar{1}\bar{1})$.[22] Apparently, $(10\bar{1}\bar{1})$ is a low energy face in InN.

Figure 4 shows the same growth mode as observed in ZnO when grown by chemical vapor deposition (CVD) on a thermally oxidized Si (111) substrate. In this case, there were no epitaxial relations with the SiO$_2$ substrate, and the crystals are observed to be randomly rotated around their growth axis with respect to one another (Fig. 4B). The width of the ZnO crystals is roughly 10 times that observed in the InN. This actually makes sense, because the thermal generation in ZnO is much smaller due to the wider bandgap. Therefore, a larger volume would be required to generate the same number of electron-hole pairs. An accurate comparison, however, would have to take into account the exact electric fields induced by the polar-charge, which calculation would require parameters that are generally not available at present for non-ferroelectric materials due to the great difficulty in measuring them. In the ZnO case, the rod continues to extend laterally at a slow pace also during its vertical growth, and for this reason, the inverted phase width is not observed to extinct at the substrate level. Another evidence for the lateral growth is the somewhat larger width at the bottom compared to the top. The pyramidal tip, observed in the InN, is missing in the ZnO, supporting our previous suggestion that this feature is not necessarily a part of this mode of growth.

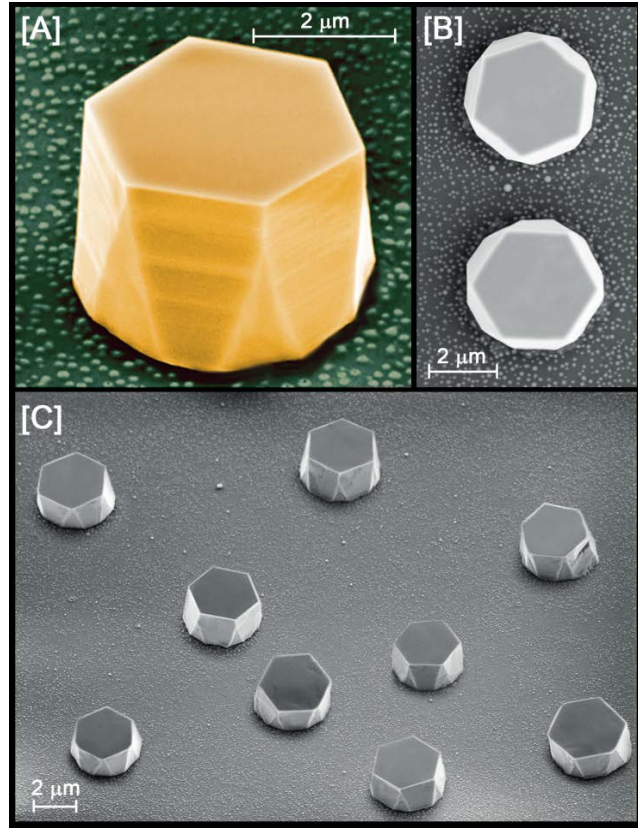

**Figure 4** - Scanning electron microscope (SEM) images of CVD-grown ZnO nano-rods on thermally oxidized Si(111): (A) false-colored close-up view showing the facet structure. (B) Head-on view showing the lack of rotational symmetry testifying the lack of epitaxial relations with the substrate. (C) Wider view showing several crystals. This image shows that the crystals grow along the c-axis which aligns perpendicular to the substrate despite the lack of epitaxial relations. The mechanism underlying this alignment is beyond the scope of this work.

Interestingly, growth of a *continuous layer* of InN rather than rods was observed, when the sapphire substrate was replaced with a heavily doped GaN template.[14] Apparently, the mobile charge exchange with the conductive substrate interfered with the proposed charging process, thus enabling the mesas to expand far enough to merge into a continuous layer.

While the proposed growth mode is limited to polar materials, it may not be limited to InN and ZnO. In the early days of GaN research, the growth of GaN directly on sapphire (without a so-called "nucleation layer") was often observed to result in what was then dubbed "hillocking". Reminiscent of our nanorods, GaN hillocking typically exhibited a different aspect ratio than that observed in our InN rods.[23] An intriguing result of the early work with GaN showed that columnar hillocks were surrounded by inversion domains.[24] Unexplained inversion domains have also been reported in the growth of c-oriented GaN nanowires on c-plane sapphire, but in all these cases the non-symmetric cross-section resulted in a non-symmetric domain appearance that rendered their explanation more difficult compared with our case.[25,26,27,28,29] Formation of fibers, such as we observed, should be limited to cases, wherein the unit cell is anisotropic and growth takes place along the polar axis. This makes the 2H polytypes of group-III nitrides a natural example. Indeed, similar AlN nanorod structures as well as inversion domains in AlN layers have been observed to grow on C-plane sapphire.[30,31] For the past twenty years, the occurrence of inversion domains in layer growth of group III nitrides has been considered a challenge to this crystal growth technology. We believe our observations and model shed new light and suggest a previously unconsidered mechanism that promotes non-isotropic growth to form crystalline fibers, when polar materials are grown on insulators. When layer growth is desired, this fiber mode of growth could be easily avoided, if the sapphire were to be replaced with, e.g., *conductive* silicon carbide.

## Materials and Methods

Growth of InN on c-plane sapphire was carried out in a hydride vapor phase epitaxy reactor. The reactor consisted of 75 mm quartz tube, placed in a three-zone horizontal furnace. One heating zone was used for hydride reaction of In with HCl gas (diluted with N2). The resulting Indium chloride was formed in an internal quartz tube (10 mm cross section) at 500 °C and was carried in its dedicated tube into the reaction zone. The reaction zone was kept at 550 °C during the process. Ammonia was delivered to the growth zone via separate 10 mm diameter quartz tube, and the InCl gas reacted with ammonia to form InN. The growth was carried out at atmospheric pressure. Ultra high purity Ar was used as a carrier gas. Typical carrier flow was 3000 sccm, HCl and $NH_3$ flows were 5 and 100 sccm, respectively.

ZnO was grown by chemical vapor deposition (CVD) in a tube furnace on a thermally oxidized Si <111> substrates. Carbo-thermally reduced ZnO was used as the Zn source (30:30 mg mixture of ZnO powder and graphite powder. The growth was carried out at 1100 ºC under a flow of 50/25/2.5 sccm of $Ar/CO_2/O_2$. The samples where situated in the quartz crucible above the Zn source. The crucible was introduced into the preheated furnace using a linear-motion feed-through for 5 min following which it was rapidly pulled out.

Cross-sections were made in a dual-beam DB235 FIB/SEM instrument (FEI Co., Hillsboro, OR, U.S.A.) at ion-beam current of 10 pA and Ga+ beam energy of 10 keV. Scanning electron microscopy was carried out in a LEO (Zeiss) 1525 FEG-SEM Field Emission Gun Scanning Electron Microscope. TEM was performed using a JEOL 2010-FEG-TEM operated at 200 keV. CBED simulation was carried out using Pierre Stadelmann's JEMS electron microscopy simulation software.[32]

## Acknowledgement

This work was funded by NSF/BSF EECS (BSF grant #2015700).

# References

1. D. Jacobsson, F. Panciera, J. Tersoff, M.C. Reuter, S. Lehmann, S. Hofmann, K.A. Dick, F.M. Ross. Interface dynamics and crystal phase switching in GaAs nanowires. *Nature* **531**, 317 (2016). DOI: 10.1038/nature17148

2. N.P. Dasgupta, J. Sun, C. Liu, S. Brittman, S.C. Andrews, J. Lim, H. Gao, R. Yan, P.D. Yang. Semiconductor Nanowires – Synthesis, Characterization, and Applications. *Adv. Matter*. **26**, 2137 (2014). DOI: 10.1002/adma.201305929

3. F.C. Frank. The influence of dislocations on crystal growth. *Disc. Faraday Soc*. **5**, 48-54 (1949). DOI: 10.1039/DF9490500048

4. R. S. Wagner and W. C. Ellis, *Appl. Phys. Lett*. **4**, 89 (1964). DOI: 10.1063/1.1753975

5. The VLS model has later been challenged, as it was further suggested that liquid may not be a strict requirement in catalytic growth and may be, in some cases, replaced by a solid phase. See article: A. I. Persson, M. W. Larsson, S. Stenstrom et al., *Nat. Mater*. 3, 677 (2004). DOI: 10.1038/nmat1220

6. J.J. Wu, S.C. Liu, Catalyst-free growth and characterization of ZnO nanorods. *J. Phys. Chem. B* **106**, 9546 (2002). DOI: 10.1021/jp025969j

7. B. Mandl, J. Stangl, E. Hilner, A.A. Zakharov, K. Hillerich, A.W. Dey, L. Samuelson, G. Bauer, K. Deppert, A. Mikkelsen. Growth Mechanism of Self-Catalyzed Group III-V Nanowires. *Nano Lett*. **10**, 4443 (2010). DOI: 10.1021/nl1022699

8. F. Glas, M.R. Ramdani, G. Patriarche, J.-C. Harmand. Predictive modeling of self-catalyzed III-V nanowire growth. *Phys. Rev. B* **88**, 195304 (2013). DOI: 10.1103/PhysRevB.88.195304

9. O. Moutanabbir, D. Isheim, H. Blumtritt, S. Senz, E. Pippel, D.N. Seidman. Colossal injection of catalyst atoms into silicon nanowires. *Nature* **496**, 78 (2013). DOI: 10.1038/nature11999

10. Kim H-M, Kim D S, Park Y S, Kim D Y, Kang T W, Chung K S, Growth of GaN nanorods by a hydride vapor phase epitaxy method. *Adv. Mater*. **14**, 991 (2002). DOI: 10.1002/1521-4095(20020705)14:13/14<991::AID-ADMA991>3.0.CO;2-L

11. K.W. Kolasinski. Catalytic growth of nanowires: Vapor–liquid–solid, vapor–solid–solid, solution–liquid–solid and solid–liquid–solid growth. *Curr. Opin. Solid State Mater. Sci.*. **10**, 182 (2006). DOI: 10.1016/j.cossms.2007.03.002

12. J. Wu, W. Walukiewicz, K.M. Yu, J.W. Ager, E.E. Haller, H. Lu, W.J. Schaff, Y. Saito, Y. Nanishi. Unusual properties of the fundamental band gap of InN. *Appl. Phys. Lett*. **80**, 3967 (2002). DOI: 10.1063/1.1482786

13. F. Qian, Y. Li, S. Gradecak, H.-G. Park, Y.J. Dong, Y. Ding, Z.L. Wang, C.M. Lieber. Multi-quantum-well nanowire heterostructures for wavelength-controlled lasers. *Nature Matter*. **7**, 701 (2008). DOI: 10.1038/nmat2253

14. I. Shalish, G. Seyogin, W.Yi, J. Bao, M.A Zimmler, E. Likovich, D.C. Bell, F. Capasso, V. Narayanamurti. Epitaxial catalyst-free growth of InN Nanorods on c-plane sapphire. *Nanoscale Res. Lett.*, **4**, 519 (2009). DOI: 10.1007/s11671-009-9276-z

15. F. Bernardini, V. Fiorentini, D. Vanderbilt. Spontaneous polarization and piezoelectric constants of III-V nitrides. *Phys. Rev B* **56**, R10024 (1997). DOI: 10.1103/PhysRevB.56.R10024

16. G. Hansdah, B.K. Sahoo. Pyroelectric effect and lattice thermal conductivity of InN/GaN heterostructures. *J. Phys. Chem.. Solids* **117**, 111 (2018). DOI: 10.1016/j.jpcs.2018.02.018

17. V.Y. Davydov, A.A. Klochikhin, R.P. Seisyan, V.V. Emtsev, S.V. Ivanov, F. Bechstedt, J. Furthmüller, H. Harima, A.V. Mudryi, J. Aderhold, O. Semchinova, J. Graul. Absorption and Emission of Hexagonal InN: Evidence of Narrow Fundamental Band Gap. *Phys. Stat. Sol. (b)* **229**, R1 (2002). DOI: 10.1002/1521-3951(200202)229:3<r1::aid-pssb99991>3.0.co;2-o

18. J.J. Haris, K.J Lee, J.B. Webb, H. Tang, I. Harrison, L.B. Flannery, T.S Cheng, C.T. Foxon. The implications of spontaneous polarization effects for carrier transport measurements in GaN. *Semicond. Sci. Technol.* **15**, 413 (2000). DOI: 10.1088/0268-1242/15/4/319

19. N. Galiana, P.P. Martin, L. Garzón, E. Rodríguez-Cañas, C. Munuera, F. Esteban-Betegón, M. Varela, C. Ocal, M. Alonso, A. Ruiz. Formation of pyramid-like nanostructures in MBE-grown Si films on Si(001). *Appl Phys A* **102**, 731 (2011). DOI: 10.1007/s00339-010-5974-8

20. A.A. Golovin, S.H. Davis, A.A. Nepomnyashchy. Model for faceting in a kinetically controlled crystal growth. *Phys. Rev. E* **59**, 803 (1999). DOI: 10.1103/PhysRevE.59.803

21. J.-N. Aqua, T. Frisch. Influence of surface energy anisotropy on the dynamics of quantum dot growth. *Phys. Rev. B* **82**, 085322 (2010). DOI: 10.1103/PhysRevB.82.085322

22. Sharma N, Thomas P, Tricker D, Humphreys C. Chemical mapping and formation of V-defects in InGaN multiple quantum wells. *Appl. Phys. Lett.* **77**, 1274 (2000). DOI: 10.1063/1.1289904

23. J. Marini, J. Leathersich, I. Mahaboob, J. Bulmer, N. Newman, F. Shahedipour-Sandvik. MOCVD growth of N-polar GaN on on-axis sapphire substrate: Impact of AlN nucleation layer on GaN surface hillock density. *J. Cryst. Grow.* **442**, 25 (2016). DOI: 10.1016/j.jcrysgro.2016.02.029

24. J.L. Rouviere, M. Arlery, R. Niebuhr, K.H. Bachem, O. Briot. Transmission electron microscopy characterization of GaN layers grown by MOCVD on sapphire. *Mater. Sci. Eng. B* **43**, 161 (1997). DOI: 10.1016/S0921-5107(96)01855-7

25. R. Koester, J.S. Hwang, C. Durand, D. Le Si Dang, J. Eymery. Self-assembled growth of catalyst-free GaN wires by metal–organic vapour phase epitaxy. *Nanotechnology* **21**, 015602 (2010). DOI: 10.1088/0957-4484/21/1/015602

26. X.J. Chen, G. Perillat-Merceroz, D. Sam-Giao, C. Durand, J. Eymery. Homoepitaxial growth of catalyst-free GaN wires on N-polar substrates. *Appl. Phys. Lett.* **97**, 151909 (2010); DOI: 10.1063/1.3497078

27. B. Alloing, S. Vézian, O. Tottereau, P. Vennéguès, E. Beraudo, J. Zuniga-Pérez. On the polarity of GaN micro- and nanowires epitaxially grown on sapphire (0001) and Si(111) substrates by metal organic vapor phase epitaxy and ammoniamolecular beam epitaxy. *Appl. Phys. Lett.* **98**, 011914 (2011); DOI: 10.1063/1.3525170

28. S. Labat, M.-I. Richard, M. Dupraz, M. Gailhanou, G. Beutier, M. Verdier, F. Mastropietro, T.W. Cornelius, T.U. Schu¨lli, J. Eymery, O. Thomas. Inversion Domain Boundaries in GaN Wires Revealed by Coherent Bragg Imaging. *ACS nano* **9**, 9210 (2015). DOI: 10.1021/acsnano.5b03857

29. Y.N. Ahn, S.H. Lee, S.K. Lim, K.J. Woo, H.n Kim. The role of inversion domain boundaries in fabricating crack-free GaNfilms on sapphire substrates by hydride vapor phase epitaxy. *Mater. Sci. .Eng. B* **193**, 105 (2015). DOI: 10.1016/j.mseb.2014.11.012

30. J. Yang, T.W. Liu, C.W. Hsu, L.C. Chen, K.H. Chen, C.C. Chen, Controlled growth of aluminum nitride nanorod arrays via chemical vapor deposition. *Nanotechnology* **17** (11): S321-S326 (2006). DOI: 10.1088/0957-4484/17/11/S15

31. V. Kueller, A. Knauer, F. Brunner, A. Mogilatenko, M. Kneissl, M. Weyers. Investigation of inversion domain formation in AlN grown on sapphire by MOVPE. *Phys. Stat. Sol. C* **9**, 496 (2012). DOI: 10.1002/pssc.201100495

32. P. Stadelmann. 2004 JEMS, electron microscopy software, java version. http://cimewww.epfl.ch/people/stadelmann/jemsWebSite/jems.html